\documentclass[prl,superscriptaddress,twocolumn,showpacs,preprintnumbers,amsmath,amssymb]{revtex4}

\usepackage{epsfig}
\usepackage{dcolumn}
\usepackage{bm}
\usepackage{ulem}

\usepackage{graphicx}

\def\simge{\mathrel{%
   \rlap{\raise 0.511ex \hbox{$>$}}{\lower 0.511ex \hbox{$\sim$}}}}
\def\simle{\mathrel{
   \rlap{\raise 0.511ex \hbox{$<$}}{\lower 0.511ex \hbox{$\sim$}}}}

\newcommand{\pt}{p_T}
\newcommand{\Et}{E_T}

\newcommand{\MET}{\mbox{$E\kern-0.60em\raise0.10ex\hbox{/}_{T}$}}
\newcommand{\vecMET}{\mbox{$\vec{E}\kern-0.60em\raise0.10ex\hbox{/}_{T}$}}
\newcommand{\METrel}{\mbox{$E\kern-0.60em\raise0.10ex\hbox{/}_{T,rel}$}}
\newcommand{\MetDeltaPhi}{\Delta\phi_{E\kern-0.50em\raise0.08ex\hbox{\scriptsize /}_{T},(\ell{\rm,jet)}}}
\newcommand{\MetDeltaPhil}{\Delta\phi_{E\kern-0.50em\raise0.08ex\hbox{\scriptsize /}_{T},(\ell)}}

\newcommand{\Z}{Z}
\newcommand{\WW}{WW}
\newcommand{\WZ}{WZ}
\newcommand{\ZZ}{ZZ}

\newcommand{\HWW}{H\rightarrow WW^{(\ast)}}

\newcommand{\ppbar}{p \overline{p}}

\newcommand{\ttbar}{t \overline{t}}

\newcommand{\mh}{m_H}

\begin{document}

\title{ \boldmath Inclusive Search for Standard Model Higgs Boson Production in 
the $\WW$ Decay Channel using the CDF II Detector}

\affiliation{Institute of Physics, Academia Sinica, Taipei, Taiwan 11529, Republic of China} 
\affiliation{Argonne National Laboratory, Argonne, Illinois 60439} 
\affiliation{University of Athens, 157 71 Athens, Greece} 
\affiliation{Institut de Fisica d'Altes Energies, Universitat Autonoma de Barcelona, E-08193, Bellaterra (Barcelona), Spain} 
\affiliation{Baylor University, Waco, Texas  76798} 
\affiliation{Istituto Nazionale di Fisica Nucleare Bologna, $^{dd}$University of Bologna, I-40127 Bologna, Italy} 
\affiliation{Brandeis University, Waltham, Massachusetts 02254} 
\affiliation{University of California, Davis, Davis, California  95616} 
\affiliation{University of California, Los Angeles, Los Angeles, California  90024} 
\affiliation{University of California, San Diego, La Jolla, California  92093} 
\affiliation{University of California, Santa Barbara, Santa Barbara, California 93106} 
\affiliation{Instituto de Fisica de Cantabria, CSIC-University of Cantabria, 39005 Santander, Spain} 
\affiliation{Carnegie Mellon University, Pittsburgh, Pennsylvania  15213} 
\affiliation{Enrico Fermi Institute, University of Chicago, Chicago, Illinois 60637}
\affiliation{Comenius University, 842 48 Bratislava, Slovakia; Institute of Experimental Physics, 040 01 Kosice, Slovakia} 
\affiliation{Joint Institute for Nuclear Research, RU-141980 Dubna, Russia} 
\affiliation{Duke University, Durham, North Carolina  27708} 
\affiliation{Fermi National Accelerator Laboratory, Batavia, Illinois 60510} 
\affiliation{University of Florida, Gainesville, Florida  32611} 
\affiliation{Laboratori Nazionali di Frascati, Istituto Nazionale di Fisica Nucleare, I-00044 Frascati, Italy} 
\affiliation{University of Geneva, CH-1211 Geneva 4, Switzerland} 
\affiliation{Glasgow University, Glasgow G12 8QQ, United Kingdom} 
\affiliation{Harvard University, Cambridge, Massachusetts 02138} 
\affiliation{Division of High Energy Physics, Department of Physics, University of Helsinki and Helsinki Institute of Physics, FIN-00014, Helsinki, Finland} 
\affiliation{University of Illinois, Urbana, Illinois 61801} 
\affiliation{The Johns Hopkins University, Baltimore, Maryland 21218} 
\affiliation{Institut f\"{u}r Experimentelle Kernphysik, Karlsruhe Institute of Technology, D-76131 Karlsruhe, Germany} 
\affiliation{Center for High Energy Physics: Kyungpook National University, Daegu 702-701, Korea; Seoul National University, Seoul 151-742, Korea; Sungkyunkwan University, Suwon 440-746, Korea; Korea Institute of Science and Technology Information, Daejeon 305-806, Korea; Chonnam National University, Gwangju 500-757, Korea; Chonbuk National University, Jeonju 561-756, Korea} 
\affiliation{Ernest Orlando Lawrence Berkeley National Laboratory, Berkeley, California 94720} 
\affiliation{University of Liverpool, Liverpool L69 7ZE, United Kingdom} 
\affiliation{University College London, London WC1E 6BT, United Kingdom} 
\affiliation{Centro de Investigaciones Energeticas Medioambientales y Tecnologicas, E-28040 Madrid, Spain} 
\affiliation{Massachusetts Institute of Technology, Cambridge, Massachusetts  02139} 
\affiliation{Institute of Particle Physics: McGill University, Montr\'{e}al, Qu\'{e}bec, Canada H3A~2T8; Simon Fraser University, Burnaby, British Columbia, Canada V5A~1S6; University of Toronto, Toronto, Ontario, Canada M5S~1A7; and TRIUMF, Vancouver, British Columbia, Canada V6T~2A3} 
\affiliation{University of Michigan, Ann Arbor, Michigan 48109} 
\affiliation{Michigan State University, East Lansing, Michigan  48824}
\affiliation{Institution for Theoretical and Experimental Physics, ITEP, Moscow 117259, Russia} 
\affiliation{University of New Mexico, Albuquerque, New Mexico 87131} 
\affiliation{Northwestern University, Evanston, Illinois  60208} 
\affiliation{The Ohio State University, Columbus, Ohio  43210} 
\affiliation{Okayama University, Okayama 700-8530, Japan} 
\affiliation{Osaka City University, Osaka 588, Japan} 
\affiliation{University of Oxford, Oxford OX1 3RH, United Kingdom} 
\affiliation{Istituto Nazionale di Fisica Nucleare, Sezione di Padova-Trento, $^{ee}$University of Padova, I-35131 Padova, Italy} 
\affiliation{LPNHE, Universite Pierre et Marie Curie/IN2P3-CNRS, UMR7585, Paris, F-75252 France} 
\affiliation{University of Pennsylvania, Philadelphia, Pennsylvania 19104}
\affiliation{Istituto Nazionale di Fisica Nucleare Pisa, $^{ff}$University of Pisa, $^{gg}$University of Siena and $^{hh}$Scuola Normale Superiore, I-56127 Pisa, Italy} 
\affiliation{University of Pittsburgh, Pittsburgh, Pennsylvania 15260} 
\affiliation{Purdue University, West Lafayette, Indiana 47907} 
\affiliation{University of Rochester, Rochester, New York 14627} 
\affiliation{The Rockefeller University, New York, New York 10021} 
\affiliation{Istituto Nazionale di Fisica Nucleare, Sezione di Roma 1, $^{ii}$Sapienza Universit\`{a} di Roma, I-00185 Roma, Italy} 

\affiliation{Rutgers University, Piscataway, New Jersey 08855} 
\affiliation{Texas A\&M University, College Station, Texas 77843} 
\affiliation{Istituto Nazionale di Fisica Nucleare Trieste/Udine, I-34100 Trieste, $^{jj}$University of Trieste/Udine, I-33100 Udine, Italy} 
\affiliation{University of Tsukuba, Tsukuba, Ibaraki 305, Japan} 
\affiliation{Tufts University, Medford, Massachusetts 02155} 
\affiliation{Waseda University, Tokyo 169, Japan} 
\affiliation{Wayne State University, Detroit, Michigan  48201} 
\affiliation{University of Wisconsin, Madison, Wisconsin 53706} 
\affiliation{Yale University, New Haven, Connecticut 06520} 
\author{T.~Aaltonen}
\affiliation{Division of High Energy Physics, Department of Physics, University of Helsinki and Helsinki Institute of Physics, FIN-00014, Helsinki, Finland}
\author{J.~Adelman}
\affiliation{Enrico Fermi Institute, University of Chicago, Chicago, Illinois 60637}
\author{B.~\'{A}lvarez~Gonz\'{a}lez$^w$}
\affiliation{Instituto de Fisica de Cantabria, CSIC-University of Cantabria, 39005 Santander, Spain}
\author{S.~Amerio$^{ee}$}
\affiliation{Istituto Nazionale di Fisica Nucleare, Sezione di Padova-Trento, $^{ee}$University of Padova, I-35131 Padova, Italy} 

\author{D.~Amidei}
\affiliation{University of Michigan, Ann Arbor, Michigan 48109}
\author{A.~Anastassov}
\affiliation{Northwestern University, Evanston, Illinois  60208}
\author{A.~Annovi}
\affiliation{Laboratori Nazionali di Frascati, Istituto Nazionale di Fisica Nucleare, I-00044 Frascati, Italy}
\author{J.~Antos}
\affiliation{Comenius University, 842 48 Bratislava, Slovakia; Institute of Experimental Physics, 040 01 Kosice, Slovakia}
\author{G.~Apollinari}
\affiliation{Fermi National Accelerator Laboratory, Batavia, Illinois 60510}
\author{J.~Appel}
\affiliation{Fermi National Accelerator Laboratory, Batavia, Illinois 60510}
% ...

\collaboration{CDF Collaboration\footnote{With visitors from $^a$University of Massachusetts Amherst, Amherst, Massachusetts 01003,
$^b$Universiteit Antwerpen, B-2610 Antwerp, Belgium, 
$^c$University of Bristol, Bristol BS8 1TL, United Kingdom,
$^d$Chinese Academy of Sciences, Beijing 100864, China, 
$^e$Istituto Nazionale di Fisica Nucleare, Sezione di Cagliari, 09042 Monserrato (Cagliari), Italy,
$^f$University of California Irvine, Irvine, CA  92697, 
$^g$University of California Santa Cruz, Santa Cruz, CA  95064, 
$^h$Cornell University, Ithaca, NY  14853, 
$^i$University of Cyprus, Nicosia CY-1678, Cyprus, 
$^j$University College Dublin, Dublin 4, Ireland,
$^k$University of Edinburgh, Edinburgh EH9 3JZ, United Kingdom, 
$^l$University of Fukui, Fukui City, Fukui Prefecture, Japan 910-0017,
$^m$Kinki University, Higashi-Osaka City, Japan 577-8502,
$^n$Universidad Iberoamericana, Mexico D.F., Mexico,
$^o$University of Iowa, Iowa City, IA  52242,
$^p$Kansas State University, Manhattan, KS 66506,
$^q$Queen Mary, University of London, London, E1 4NS, England,
$^r$University of Manchester, Manchester M13 9PL, England,
$^s$Muons, Inc., Batavia, IL 60510, 
$^t$Nagasaki Institute of Applied Science, Nagasaki, Japan, 
$^u$University of Notre Dame, Notre Dame, IN 46556,
$^v$Obninsk State University, Obninsk, Russia,
$^w$University de Oviedo, E-33007 Oviedo, Spain, 
$^x$Texas Tech University, Lubbock, TX  79609, 
$^y$IFIC(CSIC-Universitat de Valencia), 56071 Valencia, Spain,
$^z$Universidad Tecnica Federico Santa Maria, 110v Valparaiso, Chile,
$^{aa}$University of Virginia, Charlottesville, VA  22906,
$^{bb}$Bergische Universit\"at Wuppertal, 42097 Wuppertal, Germany,
$^{cc}$Yarmouk University, Irbid 211-63, Jordan,
$^{kk}$On leave from J.~Stefan Institute, Ljubljana, Slovenia, 
}}
\noaffiliation


\begin{abstract} 
We present a search for standard model (SM) Higgs boson production using 
$\ppbar$ collision data at $\sqrt{s} = 1.96$ TeV, collected with the CDF
II detector and corresponding to an integrated luminosity of $4.8\,{\rm fb^{-1}}$.
We search for Higgs bosons produced in all processes with a significant production rate and decaying
to two $W$ bosons.
We find no evidence for SM Higgs boson production and place upper limits at the 95\% confidence
level on the SM production cross section ($\sigma_{H}$) for values of the Higgs boson mass ($m_H$) 
in the range from $110$ to $200$ GeV.
These limits are the most stringent for  $m_H > 130$ GeV and are
1.29 above the predicted value of $\sigma_{H}$ for $m_H = 165$ GeV.
\end{abstract}


\maketitle

\normalsize

The standard model (SM) of particle physics unifies the 
electromagnetic and weak interactions into a single electroweak theory.  
However, experimental evidence and calculations in the framework of the SM show a 
difference of orders of magnitude in the cross section of electromagnetic and 
weak interactions at low energy. This fundamental 
difference is explained by the masses of the weak $W$ and $Z$ intermediate bosons that mediate
the weak interactions.
These massive bosons are a result of electroweak symmetry breaking, 
which in the SM occurs through the Higgs mechanism.
This theory is directly testable by the experimental observation of the Higgs boson,
which is one of the primary objectives of modern particle physics.
The production of Higgs bosons is expected to be observable at
the Tevatron~\cite{Han:1998ma} where the Higgs boson has a large enough cross section as 
calculated at next-to-next-to-leading 
order (NNLO)~\cite{Harlander:2002wh,Anastasiou:2002yz,Ravindran:2003um} for exclusions or evidence to be seen
with current datasets.

In this Letter we present a search for the production of SM Higgs bosons with 
subsequent decay to two oppositely charged $W^{(*)}$ bosons, 
where the asterisk indicates that $W$ bosons can be virtual. 
This search is most sensitive at high SM Higgs boson mass, $\mh > ~135$~GeV~\cite{units},
where the Higgs boson decay to $W$ bosons is
dominant~\cite{Djouadi:1997yw}.   
Previous published searches for high-mass SM Higgs bosons set upper limits at the 95\% confidence level (C.L.)
on the production cross section for a SM Higgs boson ($\sigma_{H}$) of at best 1.7 times
greater than the predicted value~\cite{Aaltonen:2008ec}.
The D$\O$ collaboration is concurrently reporting an updated 
search of comparable sensitivity in this channel~\cite{newd0prl}.
The results presented here improve the sensitivity of previous published searches by including new data, new
search topologies, and 
by performing an inclusive search for all SM Higgs boson production processes with significant rate: 
gluon-gluon fusion through virtual-quark 
loops (ggH)~\cite{deFlorian:2009hc,Anastasiou:2008tj}; production in association with a $W$
or $Z$ vector boson (VH)~\cite{Assamagan:2004mu,Brein:2003wg,Ciccolini:2003jy}; 
and vector boson fusion (VBF)~\cite{Assamagan:2004mu,Berger:2004pca}.  
The SM values of $\sigma_{H}$ for these processes at $m_H = 160$ GeV are 
$\sigma_{\mathrm{ggH}} = 0.439~{\mathrm{pb}} $, 
$\sigma_{\mathrm{WH}} = 0.051~{\mathrm{pb}} $, $\sigma_{\mathrm{ZH}} = 0.033~{\mathrm{pb}} $, 
and $\sigma_{\mathrm{VBF}} = 0.039~{\mathrm{pb}}$.
This inclusive
search expands the acceptance by 50\% for $m_H = 160$ GeV compared to searching for only the 
ggH production process as done previously by CDF~\cite{Aaltonen:2008ec}.


The CDF II detector consists of a solenoidal spectrometer with a silicon tracker and an open cell drift-chamber (COT)
surrounded by calorimeters 
and muon detectors~\cite{Acosta:2004yw}. The geometry is characterized using the azimuthal angle $\phi$ 
and the pseudorapidity $\eta\equiv-\ln[\tan(\theta/2)]$, where $\theta$ is the polar angle 
relative to the proton beam axis. 
Transverse energy, $\Et$, is defined to be $E\sin\theta$, where $E$ is the energy of an 
electromagnetic (EM) and hadronic calorimeter energy cluster.  
Transverse momentum, $\pt$, is the track momentum component transverse to the 
beam line.  

This analysis uses physics objects identified as jets, electrons, and muons as well as the
estimated missing transverse energy.
Electron and muon candidates (called electrons and muons for simplicity)
are typically identified using the COT and EM calorimeter or muon chambers respectively and
are described in detail below.
Jet candidates (jets) are measured using the calorimeter towers with corrections
to improve the estimated energy~\cite{Bhatti:2005ai} and are required to have a measured 
$\Et$ greater than $15$ GeV and $|\eta|<2.5$.
The missing transverse energy vector, $\vecMET$, is defined as the opposite of the vector sum of the 
$\Et$ of all calorimeter towers, corrected to produce
the correct average calorimeter response to jets and for the calorimeter response to muons.

The search is based on the requirement that events
contain two charged
leptons resulting from the decays of the final-state vector bosons.
These leptons have opposite charge except in
the case of the VH channel where they can have
the same charge.
We also make
requirements on the $\MET$ (explained below), which is indicative of the presence of
neutrinos, in opposite-charge di-lepton events.
The Higgs boson signature can also involve jets of hadrons produced from the decay of one of the vector bosons
in the VH process, forward quarks in the VBF process, or from the radiation of gluons.

One lepton must be identified by a trigger which performs
real time selection of electrons or muons.
One electron trigger 
requires an EM energy cluster in the central calorimeter ($|\eta|<1.1$) with $\Et >18$ GeV pointed to by 
a COT track with $\pt > 8$ GeV.  A second electron trigger requires an EM energy cluster with $\Et>20$ GeV in the 
forward calorimeter ($1.2<|\eta|<2.0$) and uncorrected $\MET > 15$ GeV. 
Muon triggers are based on track segments in the muon chambers that are 
matched to a COT track with $\pt > 18$ GeV. Trigger efficiencies are measured using samples of observed leptonic $Z$ 
decays~\cite{Abulencia:2005ix}.


The selected events consist primarily of background SM processes from three categories.  
The first category contains processes 
that like the signal result in two charged 
leptons, $\MET$, and possible jets, such as
two $W$ bosons contributed by direct production ($WW$), production of two $Z$ bosons
($ZZ$) where one $Z$ boson decays to neutrinos, and top-quark pair
production ($t \bar t$) where the $W$ bosons from top-quark decay subsequently
decay leptonically.  
The second category consists of processes such as 
Drell-Yan (DY) production with possible jets where the 
observed $\MET$ originates from the mis-measurement 
of lepton or jet energies.
Also in this
category are $\ZZ$ and $\WZ$ 
production where one or more of the final state charged 
leptons are unobserved.  
The third category includes $W$+jets ($W$j) and $W\gamma$ 
production where a final state jet or gamma is misidentified
as a charged lepton.

Higgs boson candidates are selected and contributions from the last two categories of background 
are reduced by applying the following initial selection.  
At least one charged lepton is required to match the lepton found in the trigger and have $\Et(\pt) > 20$ GeV for 
electrons (muons).  The second charged lepton is required to have $\Et(\pt) > 10$ GeV 
except in events with same charge leptons, where 
both leptons are required to have $\Et(\pt) > 20$ GeV.  To reduce backgrounds from processes resulting in
objects misidentified as charged leptons from vector boson decay (fake leptons), we employ a modified version 
of the lepton identification strategy developed for the $WZ$ 
observation analysis~\cite{Abulencia:2007tu,Aaltonen:2008ec}. Candidate 
leptons are separated into seven categories: two for electrons; four for muons; and one for isolated tracks that project to 
detector regions with insufficient calorimeter coverage for energy measurements. 
The electron categories are distinguished by whether the electron is found 
using the central or the forward calorimeter.  Electrons are further purified by
a likelihood selection based on track quality, track-calorimeter matching, calorimeter energy, 
calorimeter profile shape, and isolation information.
Two of the muon categories use muons found in the central or forward muon chambers 
and the other two use tracks consistent with originating from minimum 
ionizing particles in either the central or forward calorimeters.  Leptons are selected to be isolated by
requiring that the sum of the $\Et$ for the calorimeter towers (or for central muons the sum of track momenta) 
in a cone of $\Delta R = \sqrt{(\Delta\eta)^2 + (\Delta\phi)^2} < 0.4$ around 
the lepton is less than 10\% of the electron $\Et$ (muon $\pt$).  
Backgrounds involving 
mis-measured $\MET$ are reduced by requiring $\METrel > 25~\mathrm{GeV}$ 
for di-electrons, di-muons or events involving isolated tracks 
and $\METrel > 15~\mathrm{GeV}$ for electron-muon events, where 
$\METrel \equiv \MET \mbox{~if~}  \MetDeltaPhi > \frac{\pi}{2},~\MET\sin{\MetDeltaPhi} \mbox{~if~} \MetDeltaPhi < \frac{\pi}{2}$,
and $\MetDeltaPhi$ is the angle between the $\vecMET$ direction
and the nearest lepton or jet.     The $\METrel$ selection is not applied for same charge lepton events since 
mis-measured $\MET$ backgrounds are not large.
We reduce DY and heavy flavor backgrounds by requiring that the invariant mass of the 
lepton pair be greater than $16$ GeV.

We further subdivide the observed events into six analysis channels based on 
jet multiplicity, lepton categories, and lepton charge combinations.
The division is designed to optimize the sensitivity to the various 
Higgs boson production mechanisms~\cite{Hidas:2008zz}.  
Five of the channels have signatures with opposite-charge leptons. 
Events with zero jets and two 
central leptons are most sensitive to the leading order (LO) ggH process
and have $WW$ production as the dominant background. Events with
zero jets, in which one lepton is identified as a forward electron or
forward minimum ionizing track have an additional significant
background from fake lepton sources. As in the zero jet
case, we define two categories with one jet which are
additionally sensitive to VH and VBF Higgs boson production.
Events with two or more jets with any combination of
opposite-charge leptons can originate from any
Higgs boson production process and have $t\bar{t}$ as the dominant
background.
To reduce the $\ttbar$ background we reject events with $b$-quark jets~\cite{Abulencia:2006in}, 
which are identified by finding displaced vertices from tracks in the jets. 
We define a separate channel for same charge di-lepton events
with one or more jets. Here we exclude forward electrons
as the charge misidentification rate in the forward region is high.
This category consists of signal
events from VH production, where one lepton originates from the
associated vector boson decay, and Wj
events, where the jet is misidentified as a lepton.
In addition to extended acceptance for VH and VBF production the
final two channels add search topologies which are new compared to Ref.~\cite{Aaltonen:2008ec}.

The acceptances, efficiencies and kinematic properties of the 
signal and background processes are determined primarily using simulation. 
Events are simulated with the {\textsc MC@NLO} program for $\WW$  \cite{Frixione:2002ik}, {\sc pythia}
for $\HWW$, DY, $\WZ$, $\ZZ$, and $\ttbar$  \cite{Sjostrand:2006za}, and the
generator described in  Ref.~\cite{Baur:1992cd} for $W\gamma$.  The response of the CDF II detector is then estimated with 
a {\sc geant}-based simulation  \cite{GEANT}.  
The cross sections for each process are normalized to NNLO calculations 
with logarithmic resummation (ggH~\cite{deFlorian:2009hc,Anastasiou:2008tj}), 
NNLO (VH~\cite{Assamagan:2004mu,Brein:2003wg,Ciccolini:2003jy} and $\ttbar$ for a top-quark mass of
 $172.4$ GeV~\cite{Moch:2008ai}), 
and next-to-leading order calculations (VBF~\cite{Assamagan:2004mu,Berger:2004pca}, 
$\WW$~\cite{Frixione:2002ik}, $\WZ$ and $\ZZ$~\cite{Campbell:1999ah}, and $W\gamma$~\cite{Baur:1997kz}).
Efficiency corrections for the simulated CDF II detector response for lepton, 
photon conversion, and $b$-jet reconstruction and identification are determined using samples of 
observed $\Z\rightarrow\ell^{+}\ell^{-}$, photon conversions, and $b$-jets events, respectively.    
The probability that a jet will be misidentified as a lepton is measured 
using a sample of observed events collected with jet-based
triggers and corrected for the contributions of leptons from $W$ and $Z$ decays.  
These probabilities are applied to each jet in a $W$j
enriched sample to estimate the number of $W$j events that pass the selection~\cite{Hsu:2008zz}.

Based on the selection described above we expect 
      594  $\pm$   63  $\WW$,
      97 $\pm$ 13  $\WZ$ and $\ZZ$,
      196 $\pm$ 32  $t\bar{t}$,
      339  $\pm$ 61   DY, and
      404 $\pm$ 72 $W\gamma$ and $W$j events, for a total of 1630 $\pm$ 140 estimated background events. 
As an example, for a SM Higgs boson with $m_H = 160$ GeV we expect 
21.5  $\pm$   4.7  ggH,
      4.38 $\pm$ 0.57  WH,
      1.59 $\pm$ 0.21  ZH, and
      1.61 $\pm$ 0.26  VBF events, for a total of 29.1 $\pm$ 4.9 Higgs boson events.
We observe 1648 events.    The indicated uncertainties are systematic and are described below.

After the initial selection the proportion of expected signal
versus background is not sufficient to allow a significant
result to be extracted quantifying the amount of signal present.
Discrimination of signal from background is
greatly enhanced by employing multivariate techniques.  
We train neural networks (NN) using the Neurobayes~\cite{Feindt:2006pm}
program with a combination of background events and simulated signal events for each analysis channel and 
at each of 14 hypothesized $m_H$ values in the range $110 \le m_H \le 200$ GeV.
The inputs to the NNs are based on kinematic quantities
selected to exploit features such as the spin correlation between the $W$ bosons in Higgs boson decay, 
which results in the charged 
leptons from the $W$ decays tending to be more
collinear than in $\WW$ events; the presence of large $\MET$ from the neutrinos; 
the transverse mass of the Higgs boson, which can be reconstructed from the leptons' four-momenta and $\vecMET$; 
and the modest total energy of the Higgs boson decay products compared to $\ttbar$ decay~\cite{Hidas:2008zz}.
In the zero jet categories 
we additionally
classify each event by evaluating the
observed kinematic configuration
in a likelihood ratio of the signal
probability density divided by the sum of the signal
and background probability densities.
These probability densities are determined from 
LO matrix element calculations of the cross sections~\cite{Hsu:2008zz,Abulencia:2007tu}.  

\begin{figure}
\includegraphics[width=0.45\textwidth]{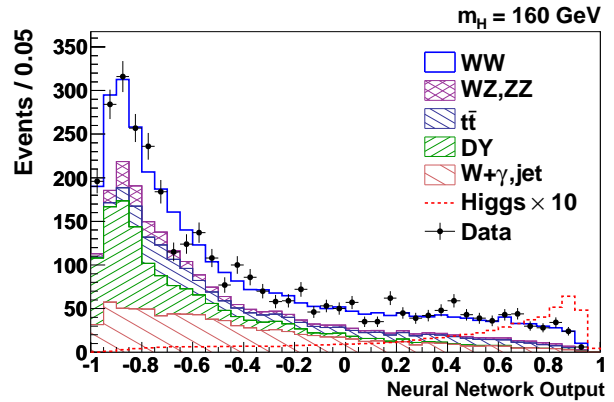}
\caption{The combined distribution of NN scores for backgrounds and a $\mh=160$ GeV Higgs boson compared
to the observed data shown with statistical uncertainties. The Higgs boson 
distribution is normalized to ten times the SM expectation.  }
\label{fig:NN}
\end{figure}

An example NN discriminant distribution for the combination of all categories is shown in Fig.~\ref{fig:NN}.  
Signal and the a-priori background
expectations for a $160$ GeV Higgs boson are shown compared to the observed
data.

We do not observe a significant excess of events 
and set upper limits at the 95\% C.L. on $\sigma_{H}$, expressed as a ratio to the expected SM rate
as a function of $m_H$.   We employ a Bayesian
technique~\cite{pdg} using a likelihood function constructed from the joint
Poisson probability of observing
the data in each bin of the discriminant NN output variables in each
channel, integrating over the
uncertainties of the normalization parameters using Gaussian priors.  A
constant prior in the signal rate is assumed.

When setting these limits we consider a variety of possible systematic effects 
including both those that change the normalization
and those that change the shape of the kinematic distributions. 
The dominant systematic uncertainties are those on
the theory predictions for the cross sections of signal and
background processes and for the data driven background estimate used for $W$j.
In addition, we consider the effect of variations from choices of renormalization and factorization scales, parton 
distribution function uncertainties, and differences between LO and higher order calculations on the acceptance of 
signal and background processes.  
The uncertainties on $\sigma_{H}$ are
5\% for WH and ZH, and 10\% for VBF.  
We estimate an additional channel-dependent
uncertainty for the ggH process of approximately $7-70$\%, to account for
scale variation cross section and acceptance uncertainties 
as a function of the number of identified jets, and a gluon PDF error of 8\% following
phenomenological NNLO studies~\cite{Anastasiou:2009bt}.  
The cross section uncertainties are 6\% for diboson production, 10\% for $\ttbar$ production, and 5\% for DY.  We estimate an 
acceptance uncertainty to account for kinematic differences between generating at LO and higher order
of 10\% for all simulated processes except $\WW$, DY, and
ggH.  We simulate the $\WW$ process at higher order
and assess a smaller uncertainty of 5\%.  The jet multiplicity and $\MET$ distributions for the DY process are
not well modeled by the simulation and we assess uncertainties from 17$-$32\% depending on channel.
We assess uncertainties of 20\% and approximately 20$-$30\%, depending on channel,
on $W\gamma$ and $W$j
backgrounds respectively, due to our modeling of conversion and fake lepton backgrounds. 
We also consider uncertainties on lepton identification and
trigger efficiencies, which range from 1.4 to 3.4\%.
Finally, we assess a 5.9\% uncertainty on the integrated luminosity.

\begin{figure}
\includegraphics[width=0.47\textwidth]{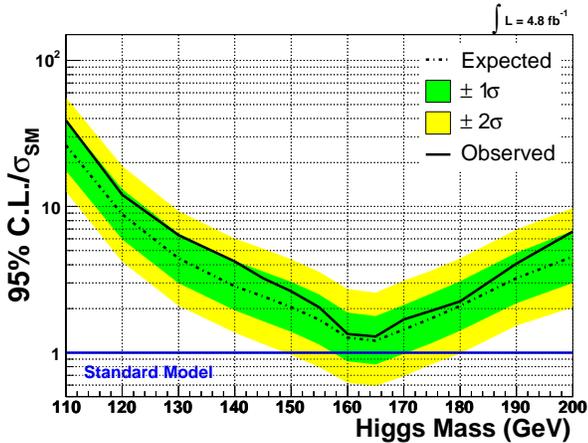}
\caption{Expected and observed upper limits at the 95\% C.L. on $\sigma_{H}$ 
presented as a ratio to the predicted SM values as a function of $\mh$. 
The dashed line represents the median expected limits, the green and yellow bands 
the estimated one and two sigma probability bands for the 
distribution of expectations, and the solid line the observed limit. }
\label{fig:limits}
\end{figure}

In Table~\ref{table:limits} and Fig.~\ref{fig:limits}
we show the median expected and observed upper limits on $\sigma_{H}$ for 14 $m_H$ 
hypotheses calculated using the techniques and
uncertainties explained above for the combination of all analysis categories. 
\begin{table}
\begin{tabular}{cccccccc}
\hline
\hline
$m_H$ (GeV)  &         110 &         120 &         130 &         140 &         145 &         150 &         155 \\
\hline
   $ \mathrm{Expected}/\sigma_{\mathrm{SM}}$ &    26.27 &     8.85 &     4.41 &     2.85 &     2.43 &     2.05  &     1.67\\ 
\hline
 $ \mathrm{Observed}/\sigma_{\mathrm{SM}}$   &    38.89 &    12.04 &     6.38 &     4.21 &     3.23 &     2.62 &     2.04 \\
\hline
\hline
$m_H$ (GeV)  &         160 &         165 &         170 &         175 &         180 &         190 &         200 \\
\hline
$ \mathrm{Expected}/\sigma_{\mathrm{SM}}$    &   1.26 &     1.20 &     1.44 &     1.72 &     2.09 &     3.24 &     4.53 \\
\hline
$ \mathrm{Observed}/\sigma_{\mathrm{SM}}$  &   1.34 &     1.29 &     1.69 &     1.94 &     2.24 &     4.06 &    6.74  \\
\hline
\hline
\end{tabular} 
\caption{Median expected and observed 95\% C.L. upper limits on $\sigma_{H}$ 
presented as a ratio to the predicted SM values of $\sigma_{H}$ as a function of $\mh$.}
\label{table:limits}
\end{table}

In conclusion, we have performed an inclusive search for SM Higgs boson
production in the two $W$ boson decay mode where the final state contains
two charged leptons.
We observe no evidence for SM Higgs boson production and set upper limits on $\sigma_{H}$.  
These limits are the most stringent to date from a single experiment for
high mass SM Higgs boson production.
We limit (at the 95\% C.L.) SM Higgs boson production to be no larger than 1.34 and 1.29
times the expected SM cross sections for $m_H = 160$ and $m_H = 165$ GeV, respectively.

We thank the Fermilab staff and the technical staffs of the participating institutions for their vital contributions. This work was supported by the U.S. Department of Energy and National Science Foundation; the Italian Istituto Nazionale di Fisica Nucleare; the Ministry of Education, Culture, Sports, Science and Technology of Japan; the Natural Sciences and Engineering Research Council of Canada; the National Science Council of the Republic of China; the Swiss National Science Foundation; the A.P. Sloan Foundation; the Bundesministerium f\"ur Bildung und Forschung, Germany; the World Class University Program, the National Research Foundation of Korea; the Science and Technology Facilities Council and the Royal Society, UK; the Institut National de Physique Nucleaire et Physique des Particules/CNRS; the Russian Foundation for Basic Research; the Ministerio de Ciencia e Innovaci\'{o}n, and Programa Consolider-Ingenio 2010, Spain; the Slovak R\&D Agency; and the Academy of Finland. 

\bibliography{document}

\begin{thebibliography}{31}
\bibitem[{uni()}]{units}
\bibinfo{note}{We use natural units where $c$ and $\hbar$ are taken as unity
  and momentum and mass are referred to in units of energy.}
\end{thebibliography}
\end{document}